# Embedded Atom Neural Network Potentials: Efficient and Accurate Machine Learning with a Physically Inspired Representation

Yaolong Zhang, Ce Hu, Bin Jiang[*]

*Hefei National Laboratory for Physical Science at the Microscale, Department of Chemical Physics, Key Laboratory of Surface and Interface Chemistry and Energy Catalysis of Anhui Higher Education Institutes, University of Science and Technology of China, Hefei, Anhui 230026, China*

[*]: corresponding author: bjiangch@ustc.edu.cn

**Abstract**

We propose a simple, but efficient and accurate machine learning (ML) model for developing high-dimensional potential energy surface. This so-called embedded atom neural network (EANN) approach is inspired by the well-known empirical embedded atom method (EAM) model used in condensed phase. It simply replaces the scalar embedded atom density in EAM with a Gaussian-type orbital based density vector, and represents the complex relationship between the embedded density vector and atomic energy by neural networks. We demonstrate that the EANN approach is equally accurate as several established ML models in representing both big molecular and extended periodic systems, yet with much fewer parameters and configurations. It is highly efficient as it implicitly contains the three-body information without an explicit sum of the conventional costly angular descriptors. With high accuracy and efficiency, EANN potentials can vastly accelerate molecular dynamics and spectroscopic simulations in complex systems at ab initio level.

*TOC graphic*

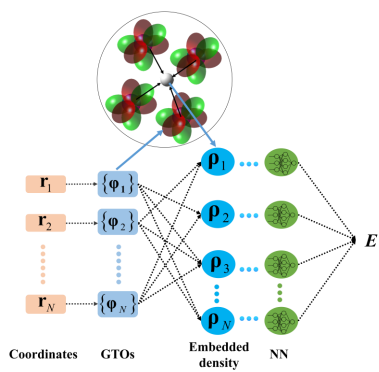

Accurate and efficient interaction potential energy surfaces (PESs) are crucial for spectroscopic, molecular dynamics, and thermodynamics simulations[1]. Traditional empirical or semi-empirical force field models such as the embedded atom method (EAM)[2-3], while physically meaningful and highly efficient, are severely limited by their accuracy. More recently, tremendous efforts have been devoted to developing machine learning (ML) based PESs[4], which are capable of representing a large set of ab initio data more accurately on chemical properties[5], molecules[6-13], gas phase and surface reactions[14-18], and materialsError! Bookmark not defined. [10, 19-22].

Since known ML techniques in computer science themselves do not recognize the intrinsic symmetry of a chemical system, it is essential to design a ML representation for a PES that preserves rotational, translational, and permutational symmetry in an accurate and efficient way[23]. In this regard, permutation invariant polynomials (PIPs) in terms of internuclear distances[24] were used as the input of neural networks (NNs)[14] and Gaussian process regression (GPR)[25] in low-dimensional systems. For high-dimensional problems, Behler and Parrinello[26] first handcrafted a set of atom centered symmetry functions (ACSFs)[23] as the input of atomistic neural networks (AtNN), which were later improved in various ways[9, 27]. More recently, the deep learning molecular dynamics (DPMD) model[10] and its symmetrized edition (DeepPot-SE)[28] map the coordinate matrix to a multi-output NN making descriptors themselves self-adapted in training. Deep tensor NN (DTNN) model[11] utilizes a vector of nuclear charges and an inter-atomic distance matrix as descriptors and its variant SchNet model[12] further introduces a continuous-filter convolutional layer to extract features from these descriptors. Alternatively, the eigenvalues of the bond distance based Coulomb matrix[8] and bi-spectrum or power spectrum of atomic densities[29] were used to measure the similarity between configurations in kernel based regression approaches, such as gradient-domain machine learning (GDML)[8] and Gaussian approximation potentials (GAP)[22], respectively. Although these ML-based potentials are much faster than ab initio calculations, but there is still room for improving the efficiency. In addition, they seem to be increasingly more intricate in their mathematical forms and more difficult to understand for chemists who are not experts in ML.

In this Letter, we propose a new NN framework born out from the EAM idea. This physically inspired embedded atom neural networks (EANN) representation is not only conceptually and numerically simple, but also very efficient and accurate, as discussed below. The EAM originates from the well-known quasi-atom[30] (or equivalently effective medium[31]) theory in which the embedding energy of an impurity in a host is a functional of the electron density of the unperturbed host without impurity, i.e., $E_i = F[\rho(\mathbf{r})]$. In general, the functional $F$ is unknown and rather complicated. Assuming that the impurity experiences a locally uniform electron density[30], the embedding energy can be approximated as a function of the scalar local electron density at the impurity site plus an electrostatic interaction[2-3]. Considering all atoms in the system as impurities embedded in the electron gas created by other atoms, in the EAM framework, the total energy of an N-atom system is just the sum over all individual impurity energies[2-3], namely,

$$E = \sum_{i=1}^{N} E_i = \sum_{i=1}^{N} \left[ F_i(\rho_i) + \frac{1}{2} \sum_{j \neq i} \phi_{ij}(r_{ij}) \right],  \qquad (1.)$$

where $F_i$ is the embedding function, $\rho_i$ is the embedded electron density at the position of atom $i$ given by the superposition of the densities of surrounding atoms, $\phi_{ij}$ is the short-range repulsive potential between atoms $i$ and $j$ depending on their distance $r_{ij}$. As the exact forms of these functions are generally unknown, they are often taken from electron gas computations[32] or fit to experimental properties with semi-empirical expressions[3]. Given these intrinsic approximations, EAM or even its modified version[33] has a limited accuracy and is mainly suitable for metallic systems.

To go beyond the EAM, we need to improve both expressions of the embedded density and the function $F$. To this end, we start from the commonly used Gaussian-type orbitals[34] (GTOs) centered at each atom,

$$\varphi_{l_x,l_y,l_z}^{\alpha,r_s}(\mathbf{r}) = x^{l_x} y^{l_y} z^{l_z} exp(-\alpha |r - r_s|^2), \qquad (2.)$$

where each atom is taken as the origin, $\mathbf{r}=(x, y, z)$ constitutes the coordinate vector of an electron, $r$ is the norm of the vector, $\alpha$ and $r_s$ are parameters that determine radial distributions of atomic orbitals, $l_x+l_y+l_z=L$ specifies the orbital angular momentum ($L$), e.g., $L=0$, 1, and 2, correspond to the $s$, $p$, and $d$ orbitals, respectively. In this representation, the embedded density of atom $i$ can be taken as the square of the linear combination of atomic orbitals from neighboring atoms, in a similar spirit as that in Hartree-Fork (HF) and density functional theory (DFT). This would generate a scalar $\rho^i$ value for the embedding atom $i$, as used in the EAM, which has been proven to offer insufficient representability for the total energy and can be improved by including the gradients of density[30]. Here, we alternatively evaluate individual electron density contribution from the same type of GTOs with the equivalent $L$, $\alpha$ and $r_s$, namely,

$$\rho_{L,\alpha,r_s}^i = \sum_{l_x,l_y,l_z}^{l_x+l_y+l_z=L} \frac{L!}{l_x!l_y!l_z!} (\sum_{j=1}^{n_{atom}} c_j \varphi_{l_x l_y l_z}^{\alpha,r_s}(\mathbf{r}_{ij}))^2, \qquad (3.)$$

where $\mathbf{r}_{ij}$ represents the Cartesian coordinates of the embedded atom *i* relative to atom *j* (or in other words the coordinates of the electron at the position of atom *i*), $c_j$ is the expansion coefficient of an orbital of atom *j*, or equivalently an element-dependent weight[27] that is optimized in the training process, the factorials of $l_x$, $l_y$, $l_z$, and $L$ constitute a pre-factor that enables the convenient transformation from Eq. (3) to an angular basis as realized by Takahashi[35], $n_{atom}$ counts the number of neighboring atoms close to the embedding atom within a sphere with a cutoff radius $(r_c)$[23]. A frequently-used cosine type cutoff function[23] is also multiplied to each orbital here to decay the interaction to zero smoothly approaching $r_c$.

Next, we replace the embedding function *F* with an atomistic NN. Since the core-core repulsion $\phi_{ij}(r_{ij})$ depends also on the internuclear distance $r_{ij}$, this part of contribution has been automatically incorporated in the NN and requires no extra terms. Eq. (1) can be thus rewritten as,

$$E = \sum_{i=1}^{N} E_i = \sum_{i=1}^{N} NN_i(\boldsymbol{\rho}^i), \qquad (4.)$$

where $\boldsymbol{\rho}^i$ consists of the orbital-dependent density components defined in Eq. (3). This naturally gives us the equivalent AtNN representation[26] in which $\boldsymbol{\rho}^i$ represents a set of local atomic density descriptors. As a result, the EANN method overcomes the intrinsic approximations on the uniform density and pairwise superposition of scalar density contributions in the EAM. It hence enables the accurate reproduction of the ab initio total energy and is no longer limited to metallic systems. The architecture of EANN is graphically illustrated in Fig. 1.

Unlike ACSFs, interestingly, these density-like descriptors are constructed in a systematical and uniform way relying on Cartesian coordinates only, with no need to classify two-body and three-body terms. It is thus beneficial to programing, especially deriving high-order analytical derivatives. More importantly, the three-body information is implicitly included by $L>0$ (*e.g. p* and *d* like) orbitals. To prove this, we define

$$f(r_{ij}) = exp(-\alpha|r_{ij} - r_s|^2). \qquad (5.)$$

According to multinomial theorem, $\rho^i_{L,\alpha,r_s}$ for the embedding atom *i* can be derived as[35],

$$\begin{aligned}
\rho^i_{L,\alpha,r_s} &= \sum_{l_x,l_y,l_z}^{l_x+l_y+l_z=L} \frac{L!}{l_x!l_y!l_z!} (\sum_{j=1}^{n_{atom}} c_j \varphi^{\alpha,r_s}_{l_x,l_y,l_z}(\mathbf{r}_{ij}))^2 \\
&= \sum_{l_x,l_y,l_z}^{l_x+l_y+l_z=L} \frac{L!}{l_x!l_y!l_z!} \sum_{j,k/=i} c_j f(r_{ij}) c_k f(r_{ik}) r_{ij}^L r_{ik}^L \frac{x_{ij}^{l_x} y_{ij}^{l_y} z_{ij}^{l_z} x_{ik}^{l_x} y_{ik}^{l_y} z_{ik}^{l_z}}{r_{ij}^L r_{ik}^L} \\
&= \sum_{j,k/=i} c_j f(r_{ij}) c_k f(r_{ik}) r_{ij}^L r_{ik}^L \sum_{l_x,l_y,l_z}^{l_x+l_y+l_z=L} \frac{L!}{l_x!l_y!l_z!} \frac{x_{ij}^{l_x} y_{ij}^{l_y} z_{ij}^{l_z} x_{ik}^{l_x} y_{ik}^{l_y} z_{ik}^{l_z}}{r_{ij}^L r_{ik}^L} \\
&= \sum_{j,k/=i} c_j f(r_{ij}) c_k f(r_{ik}) r_{ij}^L r_{ik}^L (\frac{x_{ij} x_{ik} + y_{ij} y_{ik} + z_{ij} z_{ik}}{r_{ij} r_{ik}})^L \\
&= \sum_{j,k/=i} c_j f(r_{ij}) c_k f(r_{ik}) r_{ij}^L r_{ik}^L (\cos\theta_{ijk})^L.
\end{aligned} \qquad (6.)$$

Apparently, the embedded density is actually dependent on the enclosed angle of atoms *j-i-k*, when $L>0$. However, Eq. (3) includes the pairwise summation of *j* only and does not necessarily take the costly double-loop sum over *j* and *k* as in ACSFs, giving rise to substantially lower computational cost as tested below. In addition, Eq. (6) makes the density vector invariant with respect to the overall translation and rotation, and permutation of identical atoms. In principle, our density-like descriptors form a complete set when including the orbitals up to an infinite order. In practice, we find they converge quickly and can be often truncated at most to $L=2$.

We demonstrate the representability of the EANN potentials for a number of systems ranging from small organic molecules to periodic solids, for which intensive data have been available and the details of data generation have been described elsewhere[8, 28]. In Table 1, we compare mean absolute errors (MAEs) for energy and force predictions given by a variety of successful ML-based PES models, for selected organic molecules consisting of up to 21 atoms and four elements. In general, GDML performs worse than other models because it does not scale well with respect to the number of data points due to its kernel nature, so that only 1000 structures were used for training[8]. Impressively, even reducing the number of data points to 1000, the EANN model yields comparably small errors as the GDML[8] and SchNet[12] models. Using a big data set with 50000 points, DPMD[10], DeepPot-SE[28], and SchNet models, which generally have more complicated mathematical forms and deeper NN architectures, give more accurate predictions. Remarkably, our EANN model achieves the same level of accuracy in predicting both energies and forces, with a much smaller NN architecture and a smaller number (15000~20000) of data points. For example, as seen in Table S1 in the Supplementary Information (SI), DPMD, DeepPot-SE, and SchNet models typically have several times more parameters than the EANN counterparts for fitting these molecules.

Next, Table 2 compares the performance of EANN, DPMD, and DeepPot-SE methods in representing some extended periodic systems. These systems contain more diverse phases and many atomic components, which have been argued to be more challenging for ML-based potentials, especially with conventional handcrafted descriptors[28]. For example, the PES for Pt clusters with a varying size supported on a MoS$_2$ substrate requires different structures like Pt clusters, bulk, and surfaces, and Pt/MoS$_2$ interface, all of which need to be trained in a unified model. Another challenge is the CoCrFeMnNi high entropic alloy, in which the five elements take random sites in a single lattice, yielding a great number of significantly disordered configurations. Impressively, the EANN potentials perform very well for all these challenging systems with up to 432 atoms and 5 atom types, with the prediction root mean square errors (RMSEs) in energy and force comparable to or even slightly smaller than those obtained by the DPMD and DeepPot-SE models. Again, the EANN potentials have one or two orders of magnitude fewer NN weights and biases than the DPMD and DeepPot-SE models. Moreover, only about 15%~20% randomly selected points were employed to train a unified EANN model for each system, compared to ~90% points used in the DPMD and DPMD-SE training processes. The satisfactory prediction errors indicate a promising transferability of the EANN model.

In addition to the accuracy tests, we also check the efficiency of EANN potentials. It is straightforward to compare the efficiencies of EANN and AtNN models when using the codes optimized by ourselves to the same level and keeping the number of descriptors and the NN structures identical in both models. In Table 3, taking several bulk systems with only one element type as examples (easy for comparison), we find that EANN potentials can be more than one order of magnitude faster than AtNN ones with ACSFs, due apparently to the more efficient pairwise addition nature of the orbital-based descriptors of the former. In any cases, the evaluation of descriptors represents the dominant computational bottleneck given the large number of neighboring atoms in the cutoff sphere. Meanwhile, EANN potentials yield smaller fitting errors than AtNN ones. For example, the RMSEs of energy and force for bulk Cu predicted by AtNN are 0.6 meV/atom and 137 meV/Å, respectively, much higher than those by EANN shown in Table 2. On the other hand, a direct comparison of efficiency with DPMD and SchNet models is difficult due to the distinct procedures to compute descriptors and different implementations of NNs by us and others. It should be noted that a deeper NN architecture in DPMD and SchNet models is enabled by applying residual networks (ResNets) rather than conventional feedforward NNs[10, 12]. Similarly, ResNets can be readily combined with the embedded atom density descriptors in the future to incorporate a deeper learning feature in the EANN framework.

Finally, we would like to briefly elaborate the relationships among several relevant ML models and discuss the choice of parameters. More details can be found in the SI. Our GTO-based descriptors implicitly contain the angular information via the powers of $\cos(\theta_{ijk})$, i.e., $\cos^L(\theta_{ijk})$. Three-body ACSFs[23] have the explicit form of the powers of $(1+\lambda\cos(\theta_{ijk}))$, which can not be easily transformed to an effective Cartesian representation (see the proof in the SI). Takahashi et al.[35] and BartÓk et al.[29], used a similar cosine Fourier series, or equivalently Chebyshev expansions of $\cos(\theta_{ijk})$, namely $\cos(L\theta_{ijk})$, combined with polynomials and GAPs, respectively. Artrith et al.[36] employed directly the orthonormal Chebyshev polynomials of the bond length and angle as descriptors for NNs. For the DeepPot-SE descriptors[28], we prove in the SI that the angular information is actually incorporated via the multiplication of the so-called generalized coordinate matrix resulting in only the first order of cos($\theta_{ijk}$), while the radial information is represented by the multi-output NNs rather than some Gaussian radial functions used in ACSFs. In the SchNet model[12], interestingly, a large number of radial Gaussian functions are used and the many-body character is learned by several sequential interaction blocks in the filter-generating networks. More recently, Ceriotti and coworkers[37] have derived an elegant generalized atom-density representation for the use of both kernel and NN based approaches that in form includes all descriptors mentioned above. Also discussed is its extension to fit vectorial and tensorial properties of molecules[38-39]. We consider our EANN model as one of the simplest and most efficient examples in this atom-density framework. One can of course use other types of orbitals to better capture the atomic density with more chemical insights.

One drawback of a physically inspired potential model may be the difficulty of choosing parameters in descriptors[28]. DPMD and SchNet models have been claimed to have the advantage that one only needs to specify a large size of NN descriptors or filter-generating networks to self-update the representations according to the local environment, avoiding human intervention as much as possible[12, 28]. In the EANN model, the number of orbitals depends on $L$, $\alpha$, and $r_s$. It is found that the radial distributions of adjacent orbitals have to overlap well with each other in order to have a good spatial resolution, as illustrated in Figure S1. An easy way of doing this is to define a number of evenly spacing $r_s$ grids from zero to $r_c$ and the spacing $\Delta r_s$ is related to a fixed $\alpha$, namely, $\Delta r_s = \sqrt{\beta/\alpha}$, where $\beta$ is an empirical constant of ~0.2 according to our tests. The performance of the EANN model is not very sensitive to the β value in a certain range, *e.g.* 0.1~0.3. In other words, we only need to set the maximum value of $L$ and the number of $r_s$ for each $L$ to generate a well-behaved EANN representation, making the use of the EANN model trivial. This strategy is found to work equally well for all systems studied in this work. It should be noted that Gastegger *et al.*[27] also found a similar relationship in their weighted ACSF implementation.

Summarizing, we propose a conceptually and mathematically simple, yet accurate and efficient ML model, inspired from the well-known EAM. We derive the EANN model by simply replacing the empirical expression of the scalar embedded density with a GTO-based electron density vector and the embedding functions with NNs. EANN potentials are able to achieve the same level of excellent prediction accuracy as several well-established ML models, in representing both big molecular and periodic systems, yet with much fewer parameters in NNs and fewer configurations. Because our EANN model contains the three-body information implicitly but with no need to sum over the costly angular descriptors, it is much more efficient than the conventional AtNN method using ACSFs. The parameters used to define the atomic orbitals can be readily determined with a minor human intervention. Given these desired features, the EANN model is expected to very useful for accurate and efficient molecular dynamics simulations in extended systems at ab initio level.

**Acknowledgements:** This work was supported by National Key R&D Program of China (2017YFA0303500), National Natural Science Foundation of China (21573203, 91645202, and 21722306), and Anhui Initiative in Quantum Information Technologies. We appreciate the Supercomputing Center of USTC and AM-HPC for high-performance computing services. We thank Profs. Minghui Yang and Xiao Zheng for some helpful discussions.

**Supporting Information**

Comparison of the descriptors in relevant machine learning models, training procedure, and illustration of radial distribution functions.

Table 1: Prediction MAEs by various machine learning potentials for several organic molecules. Energies and atomic forces are given in meV and meV/Å, denoted by the first and second numbers in each column. Both a small and big data set are collected, containing 1000 (1K) and 50000 (50K), or 15000~20000 (15~20K for EANN model) points, respectively. Bold numbers correspond to lowest values.

| System | Points (1K) | | | Points (15K~20K) | Points (50K) | | |
|---|---|---|---|---|---|---|---|
| | EANN | SchNet[12] | GDML[8] | EANN | SchNet[12] | DeepPot-SE[28] | DPMD[10] |
| Aspirin | 14.1, 43.0 | 16.0, 58.5 | **11.7**, **42.9** | 6.5, 18.9 | **5.2**, 14.3 | 6.7, **12.1** | 8.7, 19.1 |
| Ethanol | 4.4, 20.3 | **3.5**, **16.9** | 6.5, 34.3 | **2.1**, 4.0 | 2.2, **2.2** | 2.2, 3.1 | 2.4, 8.3 |
| Malonaldehyde | 5.9, **26.9** | **5.6**, 28.6 | 6.9, 34.7 | **3.1**, 5.1 | 3.5, **3.5** | 3.3, 4.4 | 4.0, 12.7 |
| Naphthalene | **5.0**, 11.5 | 6.9, 25.2 | 5.2, **10.0** | 4.4, 8.8 | 4.8, **4.8** | 5.2, 5.5 | **4.1**, 7.1 |
| Salicylic acid | 6.1, 22.3 | 8.7, 36.9 | **5.2**, **12.1** | 4.5, 8.2 | **4.3**, 8.2 | 5.0, **6.6** | 4.6, 10.9 |
| Toluene | **4.8**, **16.6** | 5.2, 24.7 | 5.2, 18.6 | 4.3, 9.1 | 3.9, **3.9** | 4.4, 5.8 | **3.7**, 8.5 |
| Uracil | **4.8**, 15.3 | 6.1, 24.3 | **4.8**, **10.4** | 4.3, 3.6 | 4.3, 4.8 | 4.7, **2.8** | **3.7**, 9.8 |

Table 2: Prediction RMSEs of energies (meV/atom) and atomic forces (meV/Å) by various machine learning potentials for various extended material systems. For EANN, DeepPot-SE and DPMD, 15%~20%, 90% and 90% randomly selected points used for training, respectively. Bold numbers correspond to lowest values.

| System | | EANN | | DeepPot-SE[28] | | DPMD[10] | |
|---|---|---|---|---|---|---|---|
| Bulk Cu | | **0.16** | **89** | 0.18 | 90 | 0.25 | 90 |
| Bulk Ge | | **0.07** | **31** | 0.35 | 38 | 0.60 | 35 |
| Bulk Si | | **0.08** | **28** | 0.24 | 36 | 0.51 | 31 |
| Bulk $Al_2O_3$ | | **0.11** | 51 | 0.23 | **49** | 0.48 | 55 |
| Bulk $C_5H_5N$ | Pyridine-I | **0.17** | 41 | 0.38 | **25** | 0.25 | **25** |
| | Pyridine-II | **0.35** | 49 | 0.65 | **39** | 0.43 | **39** |
| Bulk $TiO_2$ | Anatase | **1.03** | 183 | 1.78 | **181** | 3.37 | 216 |
| | Brookite | **0.55** | 96 | 0.59 | **94** | 1.97 | 109 |
| | Rutile | **0.51** | **133** | 0.96 | 137 | 1.97 | 163 |
| $MoS_2$+Pt | $MoS_2$ slab | **0.26** | **19** | 5.26 | 23 | 17.2 | 34 |
| | Pt bulk | **0.38** | **64** | 2.00 | 84 | 1.85 | 226 |
| | Pt cluster | **9.7** | **152** | 30.6 | 201 | 35.4 | 255 |
| | Pt surface | **3.8** | **86** | 6.77 | 105 | 7.12 | 187 |
| | Pt on $MoS_2$ | **1.33** | **91** | 2.62 | 94 | 5.89 | 127 |
| CoCrFeMnNi | rand1 | 2.3 | 410 | **1.68** | **394** | 6.99 | 481 |
| | rand2 | **3.3** | 415 | 5.29 | **410** | 21.7 | 576 |

Table 3: CPU time (in seconds per core) for computing energies and atomic forces of 400 configurations for three bulk metals, by AtNN with ACSFs and EANN. The number of descriptors and NN structures are kept the same. The parameter values can be found in the SI. Numbers in parentheses correspond to the time for evaluating the descriptors and its derivatives only.

| System | NN structure | Energy | | Energy+Force | |
| --- | --- | --- | --- | --- | --- |
| | | EANN | AtNN | EANN | AtNN |
| Bulk Si | 18×60×80×1 | 8.6(6.7) | 103.0(101.6) | 43.4(40.8) | 141.2(135.2) |
| Bulk Ge | 18×60×80×1 | 7.3(6.6) | 89.6(87.5) | 44.4(39.9) | 119.0(116.6) |
| Bulk Cu | 18×50×50×1 | 5.1(4.2) | 157.3(156.7) | 20.6(19.2) | 202.2(199.6) |

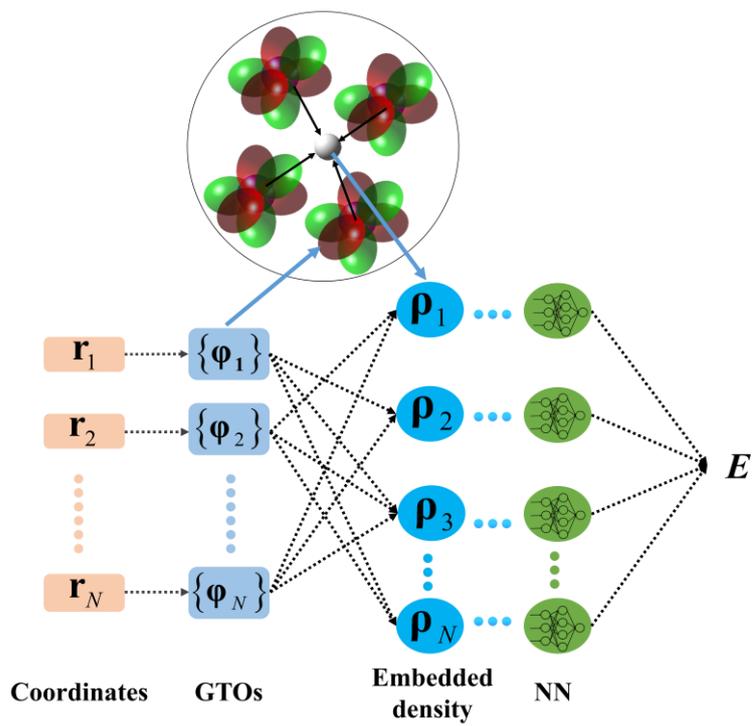

Fig. 1. Schematic architecture of the embedded atom neural network.